\documentclass[12pt]{article}
\usepackage{axodraw,bbold}

\catcode`@=12
\begin{document}
\thispagestyle{empty}
\begin{flushright} 
UCRHEP-T379\\ 
September 2004\
\end{flushright}
\vspace{0.5in}
\begin{center}
{\LARGE	\bf  Non-Abelian Discrete Family Symmetries\\  
of Leptons and Quarks\\}
\vspace{1.5in}
{\bf Ernest Ma\\}
\vspace{0.2in}
{\sl Physics Department, University of California, Riverside, 
California 92521, USA\\}
\vspace{1.5in}
\end{center}

\begin{abstract}\
Recent progress is reviewed regarding the application of the 
non-Abelian discrete symmetries $S_3$, $D_4$, and $A_4$ to the understanding 
of family structure in leptons and quarks.
\end{abstract}

\vskip 1.2in

** Talk at SI2004, Fuji-Yoshida, Japan (August 2004).

\newpage
\baselineskip 24pt

\section{Introduction}

The Standard Model (SM) has 3 families of quarks and leptons.  In the 
notation where all fermions are left-handed, they are $Q_i = (u_i,d_i)$, 
$u^c_i$, $d^c_i$; $L_i = (\nu_i,l_i)$, $l^c_i$, $[i = 1,2,3]$.  There is 
also one scalar Higgs doublet $\Phi = (\phi^+,\phi^0)$.  Under the $SU(2)_L 
\times U(1)_Y$ gauge symmetry, the allowed Yukawa couplings are $h^u_{ij} (d_i 
\phi^+ + u_i \phi^0) u^c_j$, $h^d_{ij} (d_i \bar \phi^0 - u_i \phi^-) d^c_j$, 
and $f_{ij} (l_i \bar \phi^0 - \nu_i \phi^-) l^c_j$.  Let $\langle \phi^0 
\rangle = v$, then
\begin{eqnarray}
{\cal M}_u = h^u_{ij} v &=& V_L^u \pmatrix{m_u & 0 & 0 \cr 0 & m_c & 0 \cr 
0 & 0 & m_t} (V_R^u)^\dagger, \\ 
{\cal M}_d = h^d_{ij} v^* &=& V_L^d \pmatrix{m_d & 0 & 0 \cr 0 & m_s & 0 \cr 
0 & 0 & m_b} (V_R^d)^\dagger.
\end{eqnarray}
The observed quark mixing matrix is
\begin{equation}
V_{CKM} = (V_L^u)^\dagger V_L^d \simeq \pmatrix{0.976 & 0.22 & 0.003 \cr 
-0.22 & 0.98 & 0.04 \cr 0.007 & -0.04 & 1} \simeq \pmatrix{1 & 0 & 0 \cr 
0 & 1 & 0 \cr 0 & 0 & 1}.
\end{equation}
Similarly,
\begin{eqnarray}
{\cal M}_l = f_{ij} v^* = U_L^l \pmatrix{m_e & 0 & 0 \cr 0 & m_\mu & 0 \cr 
0 & 0 & m_\tau} (U_R^l)^\dagger,
\end{eqnarray}
whereas the neutrino mass matrix is either (A) Dirac, in which case,
\begin{equation}
({\rm A}): {\cal M}_\nu^D = U_L^\nu \pmatrix{m_1 & 0 & 0 \cr 0 & m_2 & 0 \cr 
0 & 0 & m_3} (U_R^\nu)^\dagger,
\end{equation}
or (B) Majorana, in which case,
\begin{equation}
({\rm B}): {\cal M}_\nu^M = U_L^\nu \pmatrix{m_1 & 0 & 0 \cr 0 & m_2 & 0 \cr 
0 & 0 & m_3} (U_L^\nu)^T.
\end{equation}
[Note:  Neutrinos still do not have their own names!]  The observed lepton 
mixing matrix is
\begin{equation}
U_{MNS} = (U_L^l)^\dagger U_L^\nu \simeq \pmatrix{0.85 & 0.52 & 0.053 \cr 
-0.33 & 0.62 & -0.72 \cr -0.40 & 0.59 & 0.70} \simeq \pmatrix{\sqrt{2/3} & 
1/\sqrt 3 & 0 \cr -1/\sqrt 6 & 1/\sqrt 3 & -1/\sqrt 2 \cr -/\sqrt 6 & 
1/\sqrt 3 & 1/\sqrt 2}.
\end{equation}
It is clear that $U_{MNS}$ is very different from $V_{CKM}$ and the study 
of non-Abelian discrete family symmetries may help us understand why.

Since the minimal SM does not contain $\nu^c$, we should first understand 
how neutrinos get mass.  I will assume at the outset that at low energy, 
$\nu^c$ is indeed absent, in which case the only way that neutrinos can 
obtain mass is through the unique effective dimension-five operator 
first written down by Weinberg \cite{w79} 25 years ago:
\begin{equation}
{f_{ij} \over 2 \Lambda} \nu_i \nu_j \phi^0 \phi^0 \Rightarrow 
({\cal M}_\nu)_{ij} = f_{ij} {\langle \phi^0 \rangle^2 \over \Lambda}.
\end{equation}
There are 2 things to notice in the above. (1) These masses are Majorana. 
(2) These masses are necessarily ``seesaw'' in form, because $\Lambda$ is 
a large effective mass.

To understand how this operator may be realized \cite{ma98}, consider the 
6 possible pairwise products of $L_i$ and $\Phi$ in the SM, as shown below.

\begin{table}[htb]
\caption{Pairwise Products of $L_i$ and $\Phi$.}
\begin{center}
\begin{tabular}{|c|c|c|c|}
\hline
 & $SU(2)_L$ singlets & & $SU(2)_L$ triplets \\ 
\hline
(A) & $\nu_i \phi^0 - l_i \phi^+$ & (D) & $[\nu_i \phi^+, (\nu_i \phi^0 + l_i 
\phi^+)/\sqrt 2, l_i \phi^0]$ \\ 
(B) & $\nu_i l_j - l_i \nu_j$ & (E) & $[\nu_i \nu_j, (\nu_i l_j + l_i \nu_j)
/\sqrt 2, l_i l_j]$ \\ 
(C) & $\phi_1^+ \phi_2^0 - \phi_1^0 \phi_2^+$ & (F) & $[\phi^+ \phi^+, \sqrt 2 
\phi^+ \phi^0, \phi^0 \phi^0]$ \\ 
\hline
\end{tabular}
\end{center}
\end{table}

There are 4 basic combinations to obtain the effective operator of Eq.~(8). 

(I) Consider (A) $\times$ (A).  The connection must be a neutral 
fermion singlet $N$ with a large Majorana mass, i.e. the canonical seesaw 
mechanism with $m_\nu \sim \langle \phi^0 \rangle^2/m_N$.  

(II) Consider (E) $\times$ (F).  The connection is now a scalar 
triplet $(\xi^{++},\xi^+,\xi^0)$ with $m_\nu \sim \mu \langle \phi^0 \rangle^2
/m_\xi^2$, where $\mu$ is the trilinear coupling of $\xi$ to (F).  

(III) Consider (D) $\times$ (D). The connection is a Majorana 
fermion triplet $(\Sigma^+,\Sigma^0,\Sigma^-)$ with $m_\nu \sim \langle \phi^0 
\rangle^2/m_\Sigma$.  

These 3 cases are all the possible \cite{ma98} tree-level realizations of 
Eq.~(8).  (I) has dominated the literature, but (II) is getting increasingly 
more attention.

(IV) Consider (B) $\times$ (C).  This requires 2 Higgs doublets and 
a charged scalar singlet $\eta^+$ for the connection, i.e. the well-known 
one-loop radiative mechanism first proposed by Zee \cite{zee}. 

Another important question is the scale of $m_N$.  If $m_\nu \sim 1$ eV is 
assumed, then $m_N \sim 10^{13}$ GeV ($\langle \phi^0 \rangle$/$10^2$ GeV)$^2$.
However, suppose there exists a symmetry which forbids $N$ from coupling to 
$\phi^0$ but allows it to couple to $\eta^0$ instead, with $\langle \eta^0 
\rangle \sim 1$ MeV, then $m_N \sim 1$ TeV may be obtained.

Here is the simplest example \cite{ma01}.  Assign lepton number $L=0$ to $N$ 
and $(\phi^+,\phi^0)$, but $L=-1$ to $(\eta^+,\eta^0)$, then $\nu N \phi^0$ 
is forbidden, but $\nu N \eta^0$ is allowed.  Now break $L$ softly by the 
term $\mu_{12}^2 (\Phi^\dagger \eta + \eta^\dagger \Phi)$ in the Higgs 
potential, then
\begin{equation}
\langle \eta^0 \rangle \sim \mu_{12}^2 \langle \phi^0 \rangle /m_\eta^2.
\end{equation}
Let $m_\eta \sim 1$ TeV, $\langle \phi^0 \rangle \sim 10^2$ GeV, $\mu_{12}^2 
\sim 10$ GeV$^2$, then $\langle \eta^0 \rangle \sim 1$ MeV as desired.
This scenario is actually realized in the well-known case of $R-$parity 
violating supersymmetry, where $\langle \bar \eta^0 \rangle$ is $\langle 
\tilde \nu \rangle$ and $m_N$ is an effective neutralino mass.  Having 
$m_N$ at the TeV scale makes the seesaw mechanism testable at the LHC.

\section{Some non-Abelian discrete groups}

The simplest non-Abelian discrete symmetry is $S_3$, the permutation of 
3 objects, which is also the symmetry group of the equilateral triangle. 
It has 6 elements in 3 equivalence classes: $[C_1]: (123)$, $[C_2]: (231), 
(312)$, and $[C_3]: (132), (321), (213)$.  Its character table is given below.

\begin{table}[htb]
\caption{Character Table of $S_3$.}
\begin{center}
\begin{tabular}{|c|c|c|c|c|c|}
\hline
class & $n$ & $h$ & $\chi_1$ & $\chi_2$ & $\chi_3$ \\ 
\hline
$C_1$ & 1 & 1 & 1 & 1 & 2 \\
$C_2$ & 2 & 3 & 1 & 1 & $-1$ \\
$C_3$ & 3 & 2 & 1 & $-1$ & 0 \\
\hline
\end{tabular}
\end{center}
\end{table}

Here $n$ is the number of elements and $h$ is the order of each element. 
The character of each representation is its trace and must satisfy the 
following two orthogonality conditions:
\begin{eqnarray}
\sum_{C_i} n_i \chi_{ai} \chi^*_{bi} = n \delta_{ab}, ~~~~~ 
\sum_{\chi_a} n_i \chi_{ai} \chi^*_{aj} = n \delta_{ij},
\end{eqnarray}
where $n$ is the total number of elements.  The number of irreducible 
representations must be equal to the number of equivalence classes.

\begin{table}[htb]
\caption{Character Table of $D_4$.}
\begin{center}
\begin{tabular}{|c|c|c|c|c|c|c|c|}
\hline
class & $n$ & $h$ & $\chi_1$ & $\chi_2$ & $\chi_3$ & $\chi_4$ & $\chi_5$ \\ 
\hline
$C_1$ & 1 & 1 & 1 & 1 & 1 & 1 & 2 \\
$C_2$ & 1 & 2 & 1 & 1 & 1 & 1 & $-2$ \\
$C_3$ & 2 & 4 & 1 & $-1$ & $-1$ & 1 & 0 \\
$C_4$ & 2 & 2 & 1 & 1 & $-1$ & $-1$ & 0 \\
$C_5$ & 2 & 2 & 1 & $-1$ & 1 & $-1$ & 0 \\
\hline
\end{tabular}
\end{center}
\end{table}

A slightly bigger group is $D_4$, the symmetry group of the square. 
It has 8 elements in 5 equivalence classes: $[C_1]: (1234)$, $[C_2]: (3412)$, 
$[C_3]: (2341), (4123)$, $[C_4]: (1432), (3214)$, $[C_5]: (2143), (4321)$. 
Its character table is given above. 

Another interesting group is $A_4$, the even permutation of 4 objects, which 
is also the symmetry group of the tetrahedron.  It has 12 elements in 4 
equivalence classes: $[C_1]: (1234)$, $[C_2]: (1342), (4213), (2431), (3124)$, 
$[C_3]: (1423), (3241), (4132), (2314)$, $[C_4]$: (2143), (3412), (4321). 
Defining
\begin{equation}
\omega = e^{2 \pi i/3} = -{1 \over 2} + i {\sqrt 3 \over 2},
\end{equation}
its character table is given below.
 
\begin{table}[htb]
\caption{Character Table of $A_4$.}
\begin{center}
\begin{tabular}{|c|c|c|c|c|c|c|}
\hline
class & $n$ & $h$ & $\chi_1$ & $\chi_2$ & $\chi_3$ & $\chi_4$ \\ 
\hline
$C_1$ & 1 & 1 & 1 & 1 & 1 & 3 \\
$C_2$ & 4 & 3 & 1 & $\omega$ & $\omega^2$ & 0 \\
$C_3$ & 4 & 3 & 1 & $\omega^2$ & $\omega$ & 0 \\
$C_4$ & 3 & 2 & 1 & 1 & 1 & $-1$ \\
\hline
\end{tabular}
\end{center}
\end{table}

Around the year 390 BCE, the Greek mathematician Theaetetus proved that there 
were 5 and only 5 perfect geometric solids.  This posed a puzzle to Plato 
because only 4 basic elements (fire, air, water, and earth) were known, so 
a fifth element must be missing and Plato called it ``quintessence'' which 
is supposed to pervade the cosmos.  He also assigned each perfect geometric 
solid to these elements, as shown below.

\begin{table}[htb]
\caption{Five Perfect Geometric Solids.}
\begin{center}
\begin{tabular}{|c|c|c|c|c|}
\hline
solid & faces & vertices & Plato & group \\ 
\hline
tetrahedron & 4 & 4 & fire & $A_4$ \\
octahedron & 8 & 6 & air & $S_4$ \\
icosahedron & 20 & 12 & water & $A_5$ \\
hexahedron & 6 & 8 & earth & $S_4$ \\
dodecahedron & 12 & 20 & quintessence & $A_5$ \\
\hline
\end{tabular}
\end{center}
\end{table}

This was of course the first theory of everything (TOE)! In the early 19th 
century, group theory was developed and each solid was identified to have 
the symmetries indicated: $S_4$ is the group of permutation of 4 objects and 
$A_5$ is the even permutation of 5 objects.  Two pairs of solids are dual to 
one another because each can be perfectly embedded into the other and vice 
versa. The tetrahedron is special because it is self-dual.  Compare this 
to today's TOE, i.e. string theory.  There are 5 consistent string theories 
in 10 dimensions, which are related by S,T,U dualities: Type I is dual to 
Heterotic $SO(32)$, Type IIA is dual to Heterotic $E_8 \times E_8$, and 
Type IIB is self-dual.  Just as the 5 perfect geometric solids may be 
embedded in a sphere, the 5 string theories may be embedded in a single 
underlying M theory.

\section{Representations of $S_3$, $D_4$, $A_4$}

From Table 2, we see that $S_3$ has 3 irreducible representations: {\bf 1}, 
${\bf 1'}$, and {\bf 2}.  This group was used already in 1964 by Yamaguchi 
\cite{y64} for strong interactions, just before $SU(3)$ (the eightfold way) 
was proposed by Gell-Mann.  Subsequently, it was applied to the $2 \times 2$ 
quark mass matrix by Pakvasa and Sugawara \cite{ps78}. It has been studied 
by many authors, using a $2 \times 2$ real representation as follows. 
Put the vertices of the equilateral triangle in the $(x,y)$ plane at 
$1 \sim (1,0)$, $2 \sim (-1/2,\sqrt 3/2)$, $3 \sim (-1/2,-\sqrt 3/2)$. 
The 6 group elements are then represented by the corresponding matrices 
which permute these vertices.  However, there is a better way.  Let the 3 
vertices be denoted instead by $(x+iy,x-iy)$, then $1 \sim (1,1)$, 
$2 \sim (\omega,\omega^2)$, $3 \sim (\omega^2,\omega)$, where $\omega$ is 
the same quantity defined in Eq.~(11).  This ``complex'' 
representation \cite{ma91} is of course related to the real representation 
by a unitary transformation, i.e.
\begin{equation}
{1 \over \sqrt 2} \pmatrix{1 & i \cr 1 & -i} (Real) {1 \over \sqrt 2} 
\pmatrix{1 & 1 \cr -i & i} = (Complex),
\end{equation}
but is much more convenient in model building, as shown below.

\begin{table}[htb]
\caption{Real and ``Complex'' Representations of $S_3$.}
\begin{center}
\begin{tabular}{|c|c|c|c|}
\hline
class & element & real rep & complex rep \\ 
\hline
$C_1$ & (123) & $\pmatrix{1 & 0 \cr 0 & 1}$ & $\pmatrix{1 & 0 \cr 0 & 1}$ \\
\hline
$C_2$ & (231) & $\pmatrix{-1/2 & -\sqrt 3/2 \cr \sqrt 3/2 & -1/2}$ & 
$\pmatrix{\omega & 0 \cr 0 & \omega^2}$ \\
\hline
$C_2$ & (312) & $\pmatrix{-1/2 & \sqrt 3/2 \cr -\sqrt 3/2 & -1/2}$ & 
$\pmatrix{\omega^2 & 0 \cr 0 & \omega}$ \\
\hline
$C_3$ & (132) & $\pmatrix{1 & 0 \cr 0 & -1}$ & $\pmatrix{0 & 1 \cr 1 & 0}$ \\
\hline
$C_3$ & (321) & $\pmatrix{-1/2 & -\sqrt 3/2 \cr -\sqrt 3/2 & 1/2}$ & 
$\pmatrix{0 & \omega^2 \cr \omega & 0}$ \\
\hline
$C_3$ & (213) & $\pmatrix{-1/2 & \sqrt 3/2 \cr \sqrt 3/2 & 1/2}$ & 
$\pmatrix{0 & \omega \cr \omega^2 & 0}$ \\
\hline
\end{tabular}
\end{center}
\end{table}

\noindent The basic $S_3$ group multiplication rule is {\bf 2} $\times$ 
{\bf 2} = {\bf 1} + ${\bf 1'}$ + {\bf 2}, but different combinations of the 
doublet components appear in the decomposition according to which 
representation is used, as shown below.

\begin{table}[htb]
\caption{{\bf 2} $\times$ {\bf 2} Decompositions of $S_3$.}
\begin{center}
\begin{tabular}{|c|c|c|c|}
\hline
rep & {\bf 1} & ${\bf 1'}$ & {\bf 2} \\ 
\hline
real & 11+22 & $12-21$ & $\pmatrix{12+21 \cr 11-22}$ \\
\hline
complex & 12+21 & $12-21$ & $\pmatrix{22 \cr 11}$ \\
\hline
\end{tabular}
\end{center}
\end{table}

\noindent If $(\psi_1,\psi_2) \sim {\bf 2}$, then in the real representation, 
$(\psi_1^*,\psi_2^*) \sim {\bf 2}$, but in the complex representation, 
$(\psi_2^*,\psi_1^*) \sim {\bf 2}$ instead.  The invariant product of 3 
doublets in the former is given by $121+211+112-222$, whereas in the latter, 
it is simply $111+222$.

From Table 3, we see that $D_4$ has 5 irreducible representations: 
${\bf 1^{++}}$, ${\bf 1^{+-}}$, ${\bf 1^{-+}}$, ${\bf 1^{--}}$, and {\bf 2}.  
The 8 elements may be represented by the eight $2 \times 2$ matrices $\pm 1$, 
$\pm i \sigma_2$, $\pm \sigma_1$, $\pm \sigma_3$.  The basic group 
multiplication rule is
\begin{equation}
{\bf 2} \times {\bf 2} = {\bf 1^{++}} (11+22) + {\bf 1^{+-}} (12+21) + 
{\bf 1^{-+}} (11-22) + {\bf 1^{--}} (12-21).
\end{equation}
Suppose we replace $\pm \sigma_{1,3}$ by $\pm i \sigma_{1,3}$, then we obtain 
a different group of 8 elements with the same character table, except $h=4$ 
(not 2) for $C_{4,5}$ in Table 3.  This is the quaternion group $Q_8$ such 
that $(\psi_2^*,-\psi_1^*) \sim {\bf 2}$ if $(\psi_1,\psi_2) \sim {\bf 2}$, 
and
\begin{equation}
{\bf 2} \times {\bf 2} = {\bf 1^{++}} (12-21) + {\bf 1^{+-}} (11-22) + 
{\bf 1^{-+}} (12+21) + {\bf 1^{--}} (11+22).
\end{equation}

Lastly from Table 4, we see that $A_4$ has 4 irreducible representations: 
{\bf 1}, ${\bf 1'}$, ${\bf 1''}$, and {\bf 3}, with the following 
multiplication rule:
\begin{eqnarray}
{\bf 3} \times {\bf 3} &=& {\bf 1} (11+22+33) + {\bf 1'} (11+\omega^222+
\omega33) + {\bf 1''} (11+\omega22+\omega^233) \nonumber \\ && + ~
{\bf 3} (23,31,12) + {\bf 3} (32,13,21).
\end{eqnarray}

\section{$S_3$ models}

There are two recently proposed $S_3$ models of leptons and quarks. Both 
are successful phenomenologically, but they differ in their choice of 
representations and make different predictions regarding the neutrino 
mass matrix.  The first one \cite{kubo} used the conventional real 
representation and assigned 3 families of quarks, leptons (including singlet 
neutrinos $\nu^c_i$), and Higgs doublets all to {\bf 1} + {\bf 2} of $S_3$.  
Assuming first equal VEVs of the 2 Higgs doublets in {\bf 2}, there are then 
5 parameters in each Dirac mass matrix.  Adding an extra $Z_2$ symmetry such 
that $\nu^c_3$ and $\Phi_3$ are odd, while all other fields are even, 2 
parameters each in ${\cal M}^l_D$ and ${\cal M}^\nu_D$ are eliminated.  This 
leads to maximal $\nu_\mu-\nu_\tau$ mixing in an inverted hierarchy of 
neutrino masses, predicting $U_{e3} \simeq -3.4 \times 10^{-3}$.

A more recent model \cite{cfm} used the complex representation and assigned 
leptons and quarks differently from their charge conjugates.  Singlet 
neutrinos are not used.  Consider first the 2nd and 3rd families.  Assign 
$[(\nu_2,l_2),(\nu_3,l_3)] \sim {\bf 2}$, $l^c_2 \sim {\bf 1}$, $l^c_3 \sim 
{\bf 1'}$, and similarly for quarks.  Add $[\Phi_1,\Phi_2] \sim {\bf 2}$, 
then
\begin{equation}
{\cal M}_l = \pmatrix{f_1 v_2 & -f_2 v_2 \cr f_1 v_1 & f_2 v_1}, ~~~ 
{\cal M}_d = \pmatrix{h_1^d v_2 & -h_2^d v_2 \cr h_1^d v_1 & h_2^d v_1}, ~~~ 
{\cal M}_u = \pmatrix{h_1^u v_1^* & -h_2^u v_1^* \cr h_1^u v_2^* & h_2^u 
v_2^*}.
\end{equation}
Let $v_{1,2} \simeq v$, then $m_s \simeq \sqrt 2 h_1^d v$, $m_b \simeq 
\sqrt 2 h_2^d v$, $m_c \simeq \sqrt 2 h_1^u v^*$, $m_t \simeq \sqrt 2 
h_2^u v^*$, and
\begin{equation}
\theta^q_{23} \simeq {|v_2|^2 - |v_1|^2 \over |v_2|^2 + |v_1|^2} \simeq 
2 {|v_2| - |v_1| \over |v_2| + |v_1|} \simeq 0.04.
\end{equation}
Neutrino masses are assumed to be Majorana and come from heavy Higgs triplets 
$[\xi_1,\xi_2] \sim {\bf 2}$.  Thus
\begin{equation}
{\cal M}_\nu = \pmatrix{f_3 u_1 & 0 \cr 0 & f_3 u_2}.
\end{equation}
For $u_1 \neq u_2$, this implies a mismatch with ${\cal M}_l$ of
\begin{equation}
\theta^l_{23} \simeq {\pi \over 4} - {1 \over 2} \theta^q_{23} \simeq 
{\pi \over 4}.
\end{equation}
The first family is then added as a ``perturbation'': $(\nu_1,l_1), l^c_1, 
\Phi_3 \sim {\bf 1}$, and similarly for quarks.  As a result,
\begin{equation}
\theta^l_{13} \simeq {1 \over 2} \sin 2 \theta^l_{12} {\Delta m^2_{sol} \over 
\Delta m^2_{atm}},
\end{equation}
i.e. $0.008 < \theta^l_{13} < 0.032$ is predicted in a normal hierarchy of 
neutrino masses.

\section{$D_4$ models}

The symmetry $D_4$ has recently been applied \cite{grimus} to leptons 
(including $\nu^c_i$).  They are assigned as ${\bf 1^{++}}$ + {\bf 2} 
under $D_4$.  The Higgs sector has 3 doublets: $\Phi_{1,2,3} \sim 
{\bf 1^{++}}, {\bf 1^{++}}, {\bf 1^{-+}}$, and 2 singlets: $\chi_{1,2} \sim 
{\bf 2}$.  Adding an extra $Z_2$ symmetry such that $l^c_1$, $\nu^c_{1,2,3}$, 
$\Phi_1$ are odd, while all other fields are even, the charged-lepton 
mass matrix is diagonal with 3 independent eigenvalues and the neutrino 
mass matrices are given by
\begin{equation}
{\cal M}_\nu^D = \pmatrix{y_1 v_1^* & 0 & 0 \cr 0 & y_2 v_1^* & 0 \cr 0 & 
0 & y_2 v_1^*}, ~~~ 
{\cal M}_{\nu^c}^M = \pmatrix{M & y_\chi u_1 & y_\chi u_2 \cr y_\chi u_1 & 
M' & 0 \cr y_\chi u_2 & 0 & M'}. 
\end{equation}
Assuming $u_1=u_2$, we then have
\begin{equation}
{\cal M}_\nu^M = \pmatrix{x & y & y \cr y & z & w \cr y & w & z},
\end{equation}
which automatically yields \cite{ma02}
\begin{equation}
U_{MNS} = \pmatrix{\cos \theta & -\sin \theta & 0 \cr \sin \theta/\sqrt 2 & 
\cos \theta/\sqrt 2 & -1/\sqrt 2 \cr \sin \theta/\sqrt 2 & 
\cos \theta/\sqrt 2 & 1/\sqrt 2}.
\end{equation}
The neutrino masses of this model are further constrained by 
$({\cal M}^M_{\nu^c})_{23} = 0$ which implies $|({\cal M}_\nu)_{ee}| = 
m_1 m_2/m_3$, with $m_1 < m_2 < m_3$, i.e. normal ordering.  If $u_1 
\neq u_2$, then $\theta_{23} \neq \pi/4$, but $\theta^l_{13}$ remains zero.

\section{$A_4$ models}

Just as $S_3$ would be ideal to describe 2 families, $A_4$ appears perfect 
for 3 families, because it has the irreducible representations {\bf 1}, 
${\bf 1'}$, ${\bf 1''}$, and {\bf 3}.  In particular, it is a natural choice 
to have 3 degenerate neutrino masses \cite{mr01}.  Let $(\nu_i,l_i) \sim 
{\bf 3}$, $l^c_1 \sim {\bf 1}$, $l^c_2 \sim {\bf 1'}$, $l^c_3 \sim 
{\bf 1''}$, and $\Phi_i \sim {\bf 3}$, $\xi \sim {\bf 1}$, then
\begin{equation}
{\cal M}_l = \pmatrix{f_1 v_1 & f_2 v_1 & f_3 v_1 \cr f_1 v_2 & f_2 \omega 
v_2 & f_3 \omega^2 v_2 \cr f_1 v_3 & f_2 \omega^2 v_3 & f_3 \omega v_3}, ~~~~~
{\cal M}_\nu = \pmatrix{m_0 & 0 & 0 \cr 0 & m_0 & 0 \cr 0 & 0 & m_0}.
\end{equation}
For $v_1 = v_2 = v_3 = v$, ${\cal M}_l$ is diagonalized on the left by 
$U_L$ and on the right by $U_R = 1$, where
\begin{equation}
U_L = {1 \over \sqrt 3} \pmatrix{1 & 1 & 1 \cr 1 & \omega & \omega^2 \cr 
1 & \omega^2 & \omega}, ~~~~~ {\cal M}_\nu^{(e,\mu,\tau)} = U_L^\dagger 
{\cal M}_\nu U_L^* = \pmatrix{m_0 & 0 & 0 \cr 0 & 0 & m_0 \cr 0 & m_0 & 0}.
\end{equation}
In the quark sector, both ${\cal M}_{u,d}$ are diagonalized by $U_L$, hence 
$V_{CKM} = 1$ is obtained as a first approximation.

Under the most general radiative corrections, the neutrino mass matrix 
takes the form \cite{bmv}
\begin{equation}
{\cal M}_\nu = m_0 \pmatrix{1 + 2 \delta + 2 \delta' & \delta'' & \delta''^* 
\cr \delta'' & \delta & 1 + \delta \cr \delta''^* & 1 + \delta & \delta},
\end{equation}
where all complex phases have been rotated away except for that of $\delta''$. 
If $\delta''$ happens to be real, then this mass matrix is of the form of 
Eq.~(22), yielding $U_{MNS}$ of Eq.~(23) automatically.  If $\delta''$ 
has an imaginary part, then $U_{e3}$ becomes nonzero and is approximately 
given by $i Im \delta''/\sqrt 2 \delta$, thus predicting maximal $CP$ 
violation in neutrino oscillations.  The appropriate nonzero radiative 
corrections may be obtained in the context of supersymmetry \cite{hrsvv}, 
which is known to be also viable for generating a realistic $V_{CKM}$ 
\cite{bdm}.

The above $A_4$ model predicts nearly degenerate neutrino masses, with 
$\Delta m^2_{atm}$ coming from radiative corrections.  This means that 
the common mass $m_0$ of neutrinos should not be much greater than about 
0.3 eV, which is also the upper limit from neutrinoless double beta decay 
and cosmological observations.  If $m_0 < 0.1$ eV is established in 
the future, this model can be ruled out.  On the other hand, the $A_4$ 
model may be modified to accept arbitrary neutrino masses \cite{ma04}. 
Instead of using only $\xi_1 \sim {\bf 1}$, consider the addition of 
$\xi_2 \sim {\bf 1'}$, $\xi_3 \sim {\bf 1''}$, and $\xi_{4,5,6} \sim {\bf 3}$, 
with $\langle \xi^0_{1,2,3,4} \rangle \neq 0$, then
\begin{eqnarray}
{\cal M}_\nu^{(e,\mu,\tau)} &=& U_L^\dagger \pmatrix{a+b+c & 0 & 0 \cr 0 & 
a + \omega b + \omega^2 c & d \cr 0 & d & a + \omega^2 b + \omega c} U_L^* 
\nonumber \\ 
&=& \pmatrix{a+(2d/3) & d-(d/3) & c-(d/3) \cr b-(d/3) & c+(2d/3) & a-(d/3) \cr 
c-(d/3) & a-(d/3) & b+(2d/3)},
\end{eqnarray}
where $a,b,c,d$ come from $\langle \xi^0_{1,2,3,4} \rangle$ respectively. 
If $b=c$, then the eigenvalues of this mass matrix are $m_1=a-b+d$, 
$m_2=a+2b$, $m_3= -a+b+d$, and
\begin{equation}
U_{MNS} = \pmatrix{\sqrt{2/3} & 1/\sqrt 3 & 0 \cr -1/\sqrt 6 & 1/\sqrt 3 
& -1/\sqrt 2 \cr -/\sqrt 6 & 1/\sqrt 3 & 1/\sqrt 2},
\end{equation}
which is exactly the conjecture of Eq.~(7).  This predicts 
$\sin^2 2 \theta_{atm} = 1$, $U_{e3} = 0$, and $\tan^2 \theta_{sol} = 0.5$, 
for arbitrary $m_{1,2,3}$.  Experimentally, solar neutrino oscillation data 
prefer $\tan^2 \theta_{sol}$ in the vicinity of 0.4, whereas this model 
allows only the range 0.50 to 0.52 even if $b \neq c$.  On the other hand, 
if there is new physics at the TeV scale such as supersymmetry, this value 
may be shifted.

\section{Generic phenomenological consequences}

Each model discussed so far has 3 Higgs doublets.  In general, this implies 
tree-level flavor-changing charged-lepton interactions, unless ${\cal M}_l$ 
is already diagonal, as in $D_4$.  In any case, the off-diagonal Yukawa 
couplings are also suppressed by charged-lepton masses, and are usually 
safe phenomenologically.  Generically, many scalar particles should become 
observable at the LHC, decaying into various leptons.

Consider the specific case of the $A_4$ model with 3 Higgs doublets. The 
Higgs potential is given by \cite{mr01}
\begin{eqnarray}
V &=& m^2 \sum_i \Phi_i^\dagger \Phi_i + {1 \over 2} \lambda_1 ( \sum_i 
\Phi_i^\dagger \Phi_i )^2 + {1 \over 2} \lambda_2 \sum_{i,j} 
(3 \delta_{i,j} - 1) (\Phi_i^\dagger \Phi_i)(\Phi_j^\dagger \Phi_j) \nonumber 
\\ && +~ {1 \over 2} \lambda_3 \sum_{i \neq j} (\Phi_i^\dagger \Phi_j)
(\Phi_j^\dagger \Phi_i) + {1 \over 2} \lambda_4 \sum_{i \neq j} 
(\Phi_i^\dagger \Phi_j)^2.
\end{eqnarray}
It has a minimum at $v = v_1 = v_2 = v_3 = (-m^2/(3 \lambda_1 + 2 \lambda_3 + 
2 \lambda_4))^{1/2}$.  The $3 \times 3$ mass matrices of $Re \phi^0_{1,2,3}$, 
$Im \phi^0_{1,2,3}$, $\phi^\pm_{1,2,3}$ are all of the form
\begin{equation}
{\cal M}^2 = \pmatrix{a & b & b \cr b & a & b \cr b & b & a},
\end{equation}
with different values of $a,b$ in different sectors. 
Hence the linear combination $(\phi_1 + \phi_2 + \phi_3)/\sqrt 3$ behaves 
like the one Higgs doublet of the SM, and has eigenvalue $m^2 = a+2b$, 
whereas the two orthogonal combinations $\Phi',\Phi''$ are degenerate in 
mass and have eigenvalue $m^2 = a-b$.  This means that the Higgs sector 
has a residual $S_3$ symmetry, but $S_3$ is not a subgroup of $A_4$, so 
where does it come from?  The answer is that $V$ is actually also invariant 
under $S_4$.

The Yukawa couplings of $\Phi'$ and $\Phi''$ in this model are completely 
determined by the charged-lepton masses.  They are given by
\begin{eqnarray}
{\cal L}_{int} &=& {m_\tau \over v} [\overline {(\nu_e,e)}_L \Phi' + 
\overline {(\nu_\mu,\mu)}_L \Phi''] \tau_R + {m_\mu \over v} [\overline 
{(\nu_\tau,\tau)}_L \Phi' + \overline {(\nu_e,e)}_L \Phi''] \mu_R \nonumber 
\\ && +~ {m_e \over v} [\overline {(\nu_\mu,\mu)}_L \Phi' + 
\overline {(\nu_\tau,\tau)}_L \Phi''] e_R + H.c.,
\end{eqnarray}
which breaks $S_3$ explicitly.

\section{Concluding remarks}

If there is a family symmetry behind lepton (and quark) mass matrices, it is 
only evident if we know the complete Lagrangian, including the extra scalar 
(and possibly other) fields required to support this symmetry.  With our 
present incomplete knowledge, we can only assume that there is such a 
symmetry and try to find it with some educated guesses.  In the context 
of the Standard Model, the Higgs sector must be enlarged.

Texture zeros usually lead to relationships among mixing angles and mass 
ratios and they can be realized by arbitrary Abelian discrete symmetries 
\cite{zeros} supported by a possibly large number of scalars.  Non-Abelian 
discrete symmetries are very restrictive, with just a few representations. 
They may also lead to texture zeros as well as the exact equality of mass 
matrix elements which is impossible with an Abelian symmetry.  Mixing 
angles may also turn out to be unrelated to masses.  In the $D_4$ model for 
example, $\theta_{13} = 0$, $\theta_{23} = \pi/4$, and $\theta_{12}$ is 
arbitrary, independent of $m_{1,2,3}$ and $m_{e,\mu,\tau}$.

Much work has been spent on ${\cal M}_l$, ${\cal M}_\nu$, and $U_{MNS}$ 
in the past few years.  Armed with this experience, we should go back and 
tackle the old problem of ${\cal M}_u$, ${\cal M}_d$, and $V_{CKM}$ with 
hopefully a fresh perspective.

\section{Acknowledgements}

I thank M. Bando, T. Kugo, M. Tanimoto, and the other organizers of SI2004 
for their great hospitailty in Fuji-Yoshida, and the generous support 
(YITP-W-04-08) of the Yukawa Institute for Theoretical Physics at Kyoto 
University.  This work was supported in part by the U.~S.~Department of 
Energy under Grant No.~DE-FG03-94ER40837.

\end{document}